\documentclass[12pt]{article}

\begin{document}
\setcounter{page}{0}
\begin{titlepage}
\title{ Replica symmetry breaking in the transverse-field Ising
spin-glass model:Two fermionic representations}

\author{Eduardo M M Santos\\
        Alba Theumann\thanks{albath@if.ufrgs.br}\\
        Instituto de Fisica\\
        Universidade Federal do Rio Grande do Sul\\
        Av.  Bento Gon\c{c}alves 9500\\
        91501-970, Porto Alegre,RS, Brazil}

\date{}
\maketitle
\thispagestyle{empty}
\end{titlepage}

\begin{abstract}
\normalsize \noindent We analyze the infinite range Ising
spin-glass in a transverse field $\Gamma$ below the critical
temperature by a one step replica symmetry breaking theory
(1S-RSB). The set of $ n $ replicas is divided in $r $ blocks of $
m $ replicas each.We present results for different values of the
block-size parameter $ m $.The spin operators are represented by
bilinear combinations of fermionic fields and we compare the
results of two models: in the four state (4S)-model the diagonal
$S_{i}^{z}$ operator has two unphysical vanishing eigenvalues,
that are suppressed by a restraint in the two states (2S)-model.
 In the static approximation we obtain
qualitatively similar results for both models. They both exhibit a
critical temperature $T_{c}(\Gamma)$ that decreases when $\Gamma$
increases, until it reaches a quantum critical point (QCP) at the
same value of $\Gamma_{c}$ and  they are both unstable under
replica symmetry breaking in the whole spin glass phase. Below the
critical temperature we present results for the order parameters
and free energy.
 \\


\end{abstract}

\newpage
\section{Introduction}

The Ising model in a transverse field is widely studied for being the
simplest system
of interacting spins that presents a quantum critical
point(QCP)\cite{1}. We will not discuss here the extensive literature on
results for several versions
of the model, but we will concentrate instead in the transverse-field Ising
spin-glass model in the ordered region.
The experimental realizations of
this model are
 $ LiHo_{x}Y_{1-x}F_{4}$ compounds\cite{2}.

 When the transverse field $\Gamma $ vanishes, our restraint 2S- model
for two states reduces
 to the Sherrington- Kirkpatrick (SK)model of a spin glass\cite{SK}
 that in the ordered phase presents a landscape of many almost degenerate
 thermodynamic states separated by huge free energy barriers.The existence
 of these multiple states reflects itself in the breakdown of the replica
 symmetric solution of the saddle point equations below the ordering
 temperature with the onset of the Almeida-Thouless
instability\cite{ Almeida,Mezard}
 and its replacement by Parisi's replica symmetry breaking
 solution\cite{Parisi}.  The
 question is if such a picture remains true when quantum
 mechanical effects are included or if the quantum fluctuations
 will be strong enough to cause tunneling between these free
 energy barriers.

 The nature of the spin glass phase in the
transverse field Ising spin glass
model has been the subject of controversial results that seem to
depend on the representation of the spin operators,because
in the calculation of the quantum mechanical partition function
special tools are needed to deal with the non-commuting operators
entering the Hamiltonian. The method more currently used in the
study of short-range \cite{Rieger} and infinite range
\cite{Usadel1,Usadel2,Thirumalai,Buttner,Goldschmidt} spin glasses
in a transverse field is the Trotter-Suzuki formula \cite{Suzuki},
that maps a system of quantum spins in d-dimensions to a classical
system of spins in (d + 1)-dimensions, and it is suited to perform
numerical studies. In reference\cite{Thirumalai},it
was found by using the static approximation that there is a small
region in the spin glass phase where a replica symmetric (RS)
solution is stable, while in reference\cite{Buttner} it is predicted
without the use of the
static approximation that the RS solution is unstable in the whole
spin glass phase.The instability of the RS solution in the
ordered region was also obtained in reference\cite{Goldschmidt} by
analyzing a one-step replica-symmetry breaking (1S-RSB) solution,
and using only partially the static Ansatz.

 Another way of dealing with the
non-commutativity of quantum mechanical spin operators is to use
Feynman's path integral formulations \cite{Bray,Miller,Georges}
and to introduce time-ordering by means of an imaginary time
$0\leq \tau\leq\beta$, where $\beta$ is the inverse temperature. More
recently\cite{Kim} new conflicting results were obtained  by using the
formalism in ref.\cite{Bray} with the static approximation and
it was claimed that the RS solution is stable in most of the
ordered region.

A still different functional integral formulation consists in using
Grassmann variables to write a field theory with an effective
action where the spin operators in the Hamiltonian are expressed
as bilinear combinations of fermions\cite{Theumann,Oppermann}.
In a previous paper we used a replica symmetric(RS) theory with
the static approximation and we obtained
 that the RS
theory is unstable\cite{Almeida} in all the ordered
phase\cite{Theumann}. The advantage of the fermionic formulation
is that it has a natural application to problems in condensed
matter theory, where the fermion operators represent electrons
that also participate in other physical processes, like  the Kondo
effect\cite{AlbaCoqblin}.

In the present paper we  extend our previous
results\cite{Theumann} into the ordered phase with a 1S-RSB
formulation.One problem with the spin representation is that the
spin eigenstates at each site do not belong to one irreducible
representation $ S^{z}=\pm\frac{1}{2}$, but they are labeled
instead by the fermionic occupation numbers $n_{\sigma}= 0$ or 1 ,
giving two more spurious states with $S^{z}=0$. We call this the
"four states" (4S) model, and despite the presence of these two
unwanted states the 4S-Ising spin glass model describes a spin
glass transition with the same characteristics as the Sherrington
- Kirkpatrick (SK) model \cite{SK} in a replica symmetric theory.
A way to get rid of the spurious states was introduced before by
Wiethege and Sherrington\cite{Wiethege} for non-random interactions
and it consists in fixing the occupation number $n_{i\uparrow }+
n_{i\downarrow}$ by means of an integral constraint at every site.
We refer to this as the "two states" (2S)-Ising model.

In sect. 2 we analyze the 4S-Ising and 2S-Ising spin glass models
in a transverse field, within the static approximation in a one
step replica symmetric breaking (1S-RSB) theory\cite{Parisi} . The
static ansatz neglects time fluctuations and may be considered an
approximation similar to mean field theory, that describes the singularities
of the zero frequency mode\cite{Miller}.  Numerical Monte Carlo
solutions of Bray and Moore's equations indicate that the static
approximation reproduces the correct results at finite
temperatures\cite{Grempel}.

 When $\Gamma=0$ the static approximation reproduces the results
obtained by other methods, in particular for the 2S-Ising spin
glass model we recover SK equations \cite{SK}. The results in both
models are very similar; they both exhibit a critical spin glass
temperature $T_{c}(\Gamma)$ that decreases when the strength
$\Gamma$ of the transverse field increases, until it reaches a
quantum critical point(QCP) at $\Gamma _ {c}$, $T_{c}(\Gamma _
{c})=0$. The value of $\Gamma_{c}=2 \sqrt{2}J$ is the same for
both models and the 4S-Ising and 2S-Ising models are identical
close to the QCP, where $T_{c} \approx \frac{2}{3}\left(\Gamma_{c}
- \Gamma\right)$. We obtained for both models that the replica
symmetric solution is unstable \cite{Almeida} in the whole spin
glass phase, in agreement with previous results with the
Trotter-Suzuki method\cite{Goldschmidt}. In Parisi´s
theory\cite{Parisi}, the first step of RSB is achieved by dividing
$ n $ replicas in $ r $ blocks of $ m $ replicas each, and taking
the limit $ n \rightarrow 0, r \rightarrow 0 $ while keeping
finite the block parameter $ 0 \le m \le 1 $. The block parameter
$ m $ is considered to be temperature dependent and it is
determined by optimization of the free energy. In our work we
release this constraint and analyze the results for fixed values
of the block parameter $ m $.
 We present results for
the free energy, and order parameters that are consistent with
ref.\cite{Goldschmidt} but not with that of ref.\cite{Kim}.
Numerical solutions of the saddle point equations indicate that
the RSB order parameter vanishes at $T=0$. We present the model in
sect.2 and we leave sect.3 for discussions.

\newpage

\section{The model and results}

The Ising spin glass in a transverse field is represented by the
Hamiltonian

\begin{eqnarray}
 H= -\sum_{ij}J_{ij}S^{z}_{i}S^{z}_{j}-2\Gamma\sum_{i}S^{x}_{i}
\label{2.1}
\end{eqnarray}
where the sum is over the N sites of a lattice and $J_{ij}$ is a random
coupling
among all pairs of spins, with gaussian probability distribution:

\begin{equation}
P(J_{ij})=e^{-J_{ij}^{2}N/16J^{2}}{\sqrt{N/16\pi J^{2}}}\;.
\label{2.2}
\end{equation}

The spin operators are represented by auxiliary fermions fields:
\begin{eqnarray}
S^{z}_{i}=\frac{1}{2}[n_{i \uparrow}-n_{i \downarrow}]\nonumber\\
S^{x}_{i}=\frac{1}{2}[a_{i \uparrow}^{\dagger}a_{i \downarrow} +
a_{i \downarrow}^{\dagger}a_{i \uparrow}]
\label{2.3}
\end{eqnarray}
where the $a_{i\sigma}^{\dagger}(a_{i \sigma})$ are creation (destruction)
operators with
fermion anticommutatiom rules and $\sigma = \uparrow$ or$ \downarrow$
indicates the
spin projections. The number operators
$n_{i \sigma}=a_{i\sigma}^{\dagger}a_{i\sigma}= 0 or 1 $, then
 $S_{i}^{z}$ in Eq.(\ref{2.3}) has two eigenvalues $\pm\frac{1}{2}$
corresponding to $n_{i\downarrow}= 1 - n_{i\uparrow}$, and two vanishing
eigenvalues
when $n_{i\downarrow}=n_{i\uparrow}$.

We shall use the Lagrangian path integral formulation in terms of
anticommuting Grassmann fields described  in a previous
publication\cite{Theumann}, so we avoid  giving repetitious
details. We consider two models: the unrestrained, four states
model and also the two states model of Wiethege and Sherrington where
the number operators satisfy the restraint
$n_{i\uparrow}+n_{i\downarrow}=1$, what gives
$S_{i}^{z}=\pm\frac{1}{2}$, at every site \cite{Wiethege}

By using the integral representation for the  Kronecker $\delta$-function:
\begin{eqnarray}
\delta(n_{j\uparrow}+n_{j\downarrow}-1)=\frac{1}{2\pi}
\displaystyle
\int_{0}^{2\pi}dx_{j}e^{ix_{j}[n_{j\uparrow}+n_{j\downarrow}-1]}
\label{2.4}
\end{eqnarray}
we can express $Z_{4S}$ and $Z_{2S}$, the partition function for
the two models, in the compact functional integral
form\cite{Theumann}
\begin{eqnarray}
Z\{\mu\}= \int D(\varphi{\ast}\varphi)\prod_{j}\frac{1}{2\pi}
\displaystyle \int_{0}^{2\pi}dx_{j}e^{-\mu_{j}}e^{A\{\mu\}}
\label{2.5}
\end {eqnarray}
where:
\begin{eqnarray}
A\{\mu\}&=&\int_{0}^{\beta} \{\displaystyle\sum_{j,\sigma}[\varphi_{j\sigma}^{\ast}(\tau)
\frac{d}{d\tau}\varphi_{j\sigma}(\tau)+ \mu_{j} \varphi_{j\sigma}^{\ast}(\tau)
\varphi_{j\sigma}(\tau)]-\nonumber \\
        & & H(\varphi_{j \sigma}^{\ast}(\tau),\varphi_{j \sigma}(\tau)\}
\label{2.6}
\end{eqnarray}
and $\mu_{j}=0$ for the 4S-model while $\mu_{j}=ix_{j}$ for the 2S-model.
 We now follow standard procedures to get the configurational averaged
free energy per site by using the replica formalism:
\begin{equation}
F=-\frac{1}{\beta N}\lim_{n\rightarrow 0}\frac{Z(n)-1}{n}
\label{2.7}
\end{equation}
where the configurational averaged, replicated, partition function
$< Z^{n}>_{c,a}=Z(n)$ becomes, after averaging over $J_{ij}$:

\begin{eqnarray}
Z(n)&=& \int_{-\infty}^{\infty}\prod_{\alpha \beta} dq_{\alpha
\beta} e^{-N \frac{\beta^{2}J^{2}}
{2}\displaystyle \sum_{\alpha \beta}q_{\alpha \beta}^{2}} \nonumber \\
    & &\prod_{j}\{\displaystyle \prod_{\alpha}\frac{1}{2\pi}\displaystyle
\int_{0}^{2\pi}dx_{j \alpha} e^{-\mu_{j
\alpha}}\Lambda_{j}({q_{\alpha \beta}}) \}
\label{2.8}
\end{eqnarray}
with the replica index $\alpha=1,2,..,n$, and
\begin{eqnarray}
\Lambda_{j}({q_{\alpha \beta}})&=& \int D(\varphi
_{\alpha}^{\dagger} \varphi_{\alpha})exp[\sum_{\alpha}A_{j \alpha \Gamma} + \nonumber \\
& & \beta^{2}J^{2}4\displaystyle \sum_{\alpha \beta}q_{\alpha
\beta} S_{j}^{z \alpha}S_{j}^{z \beta}]
\label{2.9}
\end{eqnarray}
 We have introduced the spinors in the Fourier
representation:
\begin{eqnarray}
\underline{\psi_{i}}(\omega)=\left(\begin{array}{c}\varphi_{i\uparrow}(\omega)\\
 \varphi_{i \downarrow}(\omega)\end{array}\right)
\label{2.10}
\end{eqnarray}
and the Pauli matrices to write the spin glass part of the action
in the static approximation,where
\begin{equation}
S_{i}^{z}= \frac{1}{2} \displaystyle \sum_{\omega}\underline{\psi}^{\dagger}(\omega)
\underline{\sigma}^{z}\underline{\psi}(\omega)
\label{2.11}
\end{equation}

and by introducing the inverse propagator
\begin {equation}
\underline{\gamma}_{j}^{-1}= i\omega +\mu_{j} +\Gamma
\underline{\sigma}_{x}
\label{2.12}
\end{equation}
 we can write $A_{j \alpha \Gamma}$
\begin{equation}
 A_{j \alpha \Gamma} =
\sum_{\omega}\underline{\psi}^{\dagger \alpha} (\omega)
\underline{\gamma}_{j}^{-1}(\omega)\underline{\psi}^{\alpha}(\omega)
\label{2.13}
\end{equation}

 We assume a one step replica symmetry breaking (1S-RSB) solution\cite{Parisi}
 of the saddle point equations by separating the $n$ replicas in $r$ groups
of $m$ replicas each :
\begin{eqnarray}
q_{\alpha \alpha}= q+\bar{\chi},\nonumber \\
&  q_{\alpha \beta}=q + \delta \hskip 2 cm
I\left(\frac{\alpha}{m}\right)=
I\left(\frac{\beta}{m}\right),\nonumber \\
& q_{\alpha \beta}=q  \hskip 2 cm
I\left(\frac{\alpha}{m}\right)\neq I\left(\frac{\beta}{m}\right),
\label{2.14}
\end{eqnarray}

where q is the spin glass order parameter, $\bar{\chi}$ is related
to the static susceptibility by $\bar{\chi}=T\chi$ and
$ \delta $ is the RSB parameter. The notation $I\left(K\right) $ means
integer part.

The sums over $\alpha $ in the spin part of the action produce
again quadratic terms that can be linearized by introducing new
auxiliary fields, what makes the integration over Grassmann
variables straightforward\cite{Theumann}.  In the limit $n
\rightarrow 0, N \rightarrow \infty $ we obtain the result for the
free energy per site for the model with $ 2(p+1)$ states, $ p=0 $
or $ 1 $
\begin{equation}
\beta F_{p}=\frac{1}{2}(\beta J )^{2}\{[\bar{\chi}+ q]^2-
[\delta+q]^2 +m[{\delta}^2 +2\delta q]\}- \frac{1}{m}
\int_{-\infty}^{\infty}Dz_0 \log { \int_{-\infty}^{\infty} Dz_1
[2K_{p}(q,\bar{\chi},\delta,z)]^m}
\label{2.15}
\end{equation}
where we introduced  the standard notation $
Dy=e^{-\frac{y^2}{2}}\frac{dy}{\sqrt{2\pi}} $ and

\begin{equation}
K_{p}(q,\bar{\chi},z)=p +\int_{-\infty}^{\infty}D \xi \cosh
\sqrt{\Delta}
\label{2.16}
\end{equation}
with
\begin{equation}
h= J\sqrt{2q}z_0+
J\sqrt{2(\bar{\chi}-\delta)}\xi+ J\sqrt{2\delta}z_{1} ,
\label{2.17}
\end{equation}
\begin{equation}
\Delta= (\beta h )^{2} +(\beta \Gamma)^{2} \label{2.18}
\end{equation}
The saddle point equations for the order parameters are:
\begin{equation}
\overline{ \chi }_{p}= \int_{- \infty }^{ \infty } Dz_0
\frac{1}{\int_{-\infty}^{\infty}Dz_1 K_{p}^{m}} \displaystyle
\int_{-\infty}^{\infty}Dz_1 K_{p}^{m-1}  I_{\overline{\chi}}-
q_{p}
\label{2.19}
\end{equation}
\begin{equation}
q_{p}=\int_{-\infty}^{\infty}Dz_0
\frac{1}{\{\int_{-\infty}^{\infty}Dz_1K_{p}^{m}\}^2} \displaystyle
\{\int_{-\infty}^{\infty}Dz_1K_{p}^{m-1}\int_{-\infty}^{\infty}D
\xi \frac {\beta h}{\sqrt{\Delta}}
 \sinh{ \sqrt{ \Delta }} \}^{2}
\label{2.20}
\end{equation}
\begin{equation}
\delta_{p}=\int_{-\infty}^{\infty}Dz_0
\frac{1}{\int_{-\infty}^{\infty}Dz_1K_{p}^{m}} \displaystyle
\int_{-\infty}^{\infty}Dz_1K_{p}^{m-2}\{\int_{-\infty}^{\infty}D
\xi \frac {\beta h}{\sqrt{\Delta}} \sinh{ \sqrt{ \Delta }}
\}^{2}-q_{p} \label{2.21}
\end{equation}

where $h(\xi,z)$ is given in Eq.(\ref{2.17}) and we called:
\begin{equation}
I_{\overline{\chi}}= \int_{-\infty}^{\infty}D\xi\{ \frac{ (\beta
h)^{2} }{ \Delta } \cosh{ \sqrt{ \Delta }}+  \beta^{2} \Gamma^{2}
\frac{\sinh{\sqrt{\Delta}}}{ \Delta^{ \frac{3}{2}}} \}
\label{2.22}
 \end{equation}
 We obtain for the de Almeida-Thouless eigenvalue \cite{Almeida} and
entropy in both models:
\begin{eqnarray}
\lambda_{p}^{AT}=1-2(\beta J)^{2} \int_{-\infty }^{ \infty }
Dz_0\{
\frac{\int_{-\infty}^{\infty}Dz_1K_{p}^{m-1}I_{\overline{\chi}}}
{\int_{-\infty}^{\infty}Dz_1K_{p}^{m}}   - \nonumber\\
(\frac{\int_{-\infty}^{\infty}Dz_1K_{p}^{m-1}\int_{-\infty}^{\infty}D\xi
\sinh{\sqrt{\Delta}}\frac{\beta h}{\sqrt{\Delta}}}
{\int_{-\infty}^{\infty}Dz_1K_{p}^{m}})^2 \}^2
 \label{2.23}
\end{eqnarray}
\begin{eqnarray}
\frac{S}{k}=- \frac{3}{2}(\beta J)^{2}\{[\bar{\chi}+ q]^2-
[\delta+q]^2 +m[{\delta}^2 +2\delta q]\}+ \frac{1}{m}
\int_{-\infty}^{\infty}Dz_0 \log { \int_{-\infty}^{\infty} Dz_1
K_{p}^{m}}\nonumber\\
-(\beta \Gamma)^{2} \int_{-\infty}^{\infty}Dz_0
\frac{1}{\int_{-\infty}^{\infty}Dz_1 K_{p}^{m}}
\int_{-\infty}^{\infty}Dz_1 K_{p}^{m-1}\int_{-\infty}^{\infty}D\xi
\frac{\sinh{\sqrt\Delta}}{\sqrt{\Delta}}\nonumber\\
\label{2.24}
\end{eqnarray}
The first terms of the Landau expansion of the free energy in
powers of $q$ and $\delta$ gives:
\begin{equation}
\beta F_{p}= \beta F_{p}^{0}-\frac{1}{2}Dmq^{2}-
\frac{1}{2}D(1-m)[q+ \delta]^{2}
\label{2.25}
\end{equation}
where the coefficient D is:
\begin{eqnarray}
D=(\beta J)^2\{1-2(\beta
J)^2(\frac{I_{\overline{\chi}}}{K_{p}})^2\}
\label{2.26}
\end{eqnarray}
and the integrals are to be taken when $q= \delta= 0$. Here we
consider $0<m<1$ a free parameter and the equations will be solved
for different values of $m$. We observe that when $m=1$ we recover
the results without RSB and $q$ being the SG order parameter,
while for $m=0$ we recover also the RS result, but with $q +
\delta$ as order parameter in agreement with ref.\cite{Parisi}.
The critical line is given
by the solution of $D=0$.

The numerical results for the critical temperature $T_{c}(\Gamma)$
 were shown in ref.\cite{Theumann}.
 For large values of the transverse field $ \Gamma $ the 2S-model
and 4S-model are undistinguishable.
 When $ \Gamma = 0$, the equations (\ref{2.15})
 for the free energy of the 2S-model
($p=0$) reproduce Parisi's results for the SK model \cite{Parisi}.
 For both models our equations coincide with ref.\cite{Goldschmidt}
except for eq.(\ref{2.19}), because we find in our formalism that
the static approximation is  self-consistent. Finally, we comment
on the de Almeida-Thouless instability. The numerical solution for
$\lambda^{AT}$ in Eq.(\ref{2.23}) confirms that $\lambda^{AT}=0$
on the critical line $T_{c}(\Gamma)$ and $\lambda^{AT}<0$ for
$T<T_{c}(\Gamma)$, indicating the instability of the replica
symmetric solution in the whole ordered phase for both models and
contradicting recent results in ref.\cite{Kim}.We show in fig.1
the free energy and entropy for the 2S model($p=0$) with RSB and $
m= 0.8 , \Gamma = 0.5J $. In fig.2  we compare the free energy for
$ \Gamma= 0.5J$ in the 2S model with for different values of the
block parameter $ m $. In fig.3 we present the results for the
order parameters $ q $ and $ \delta $ for different values of $ m
$. In fig.4 we present the results for the order parameters $ q $
and $\delta $ as a function of temperature for both models in the
case of RSB, with $\Gamma = 0.5J$  and $ m=0.8 $.
\newpage
\section{Discussion}
We apply  a 1S-RSB theory\cite{Parisi} to extend to low
temperatures our previous study\cite{Theumann} of two quantum
Ising spin glass models in a transverse field by means of a path
integral formalism where the spin operators are represented by
bilinear combinations of fermionic fields, in the static
approximation. The  $n$ replicas were  separated in $ r $ groups
of $ m $ replicas each and we analyze the results for fixed values
of the block size parameter $ 0\le m \le 1 $. This differs from
Parisi's theory\cite{Parisi} where $ m $ is considered to be
temperature dependent and it is determined by optimization of the
free energy. In the unrestricted four-states 4S-model the diagonal
$S_{i}^{z}$-operator has two eigenvalues $S_{i}^{z}= \pm
\frac{1}{2}$ and two vanishing eigenvalues, while in the 2S-model
the vanishing eigenvalues are suppressed by means of an integral
constraint\cite{Wiethege}.
 Regarding the de Almeida-Thouless
instability\cite{Almeida}, we obtained before\cite{Theumann} that
the replica symmetric solution is unstable in the whole spin glass
phase for both models, contradicting the results in ref.\cite{Kim}
but in agreement with the results in ref.\cite{Buttner}and in
ref.\cite{Goldschmidt}, obtained with the method of Trotter and
Suzuki\cite{Suzuki}. The  results we present here for the
solutions of the saddle point equations,  shown in fig.3 and fig.4,
seem to indicate  that the RSB order parameter $ \delta$
vanishes at $ T=0 $ for all values of the block parameter $ m $.
The free energy ,shown in fig.2, is maximized for smaller values of
$m$  and this is consistent with the results of ref.\cite{Parisi} where it
is obtained $ m = 0$ at $ T=0 $. However, if we introduce
the value $ \delta = 0 $ in our analytic expression in eq.(\ref{2.24})
for the entropy we would recover the RS result, in discrepancy
with the numerical results of ref.\cite{Parisi} at $ T=0 $.
We believe the origin of this discrepancy lies in keeping
the block parameter $ m $
fixed.We notice in fig.3 that for $ m=0.2 $ and $ T\ge 0.2J $
the order parameters $q $ and $ \delta $ cross, that is $ \delta \ge q $.
We may conjecture thar for $ m=0 $ this behaviour extends to all
temperatures $ T \le T_{c} $, in agreement with the numerical
results\cite{Parisi}.\\
To summarize, we study the ordered phase for the Ising spin glass model
in the presence of a transverse field by using a fermionic representation
of the spin operators.The RS theory is unstable for $ T \le T_{c} $
\cite{Theumann}
and we study a 1S-RSB theory where we hold fixed the block-size parameter
$ m $ . We derive saddle point equations for the order parameters that agree
with previous results obtained with other methods\cite{Goldschmidt}.
Our theory has the advantage of computational simplicity and permit us
to clarify the importance of $ m $, but it does not describe RSB at
$ T=0 $ for finite values of $ m \ge 0.2 $. We conjecture that RSB is
restored at $ T=0 $ only for $ m=0 $.

\section{Acknowledgement}
Eduardo M M Santos acknowledges financial support from the Conselho
Nacional de Desenvolvimento Cient\'{\i}fico e Tecnol\'ogico
(CNPq).\\
\newpage

\section{Figure Captions}

fig.1 Free energy and entropy for the 2S-model with RSB,
$m= 0.8 $ ,$ \Gamma = 0.5J $
as a function of temperature\\
\\

fig.2 Free energy for the 2S-model with RSB and $ m=0.2 $(full
line), $ m =0.4 $ (squares) and $ m=0.6 $ (triangles),
as a function of temperature for $\Gamma= 0.5J $.\\
\\
fig.3 Spin glass order parameter $ q $ and RSB order parameter $
\delta $ for $ \Gamma=0.5J $ in the 2S-model.$ m=0.2 $ (full
line),$ m=0.4 $ (squares) and $ m=0.6 $ (triangles).\\
\\

fig.4 Order parameters for $m=0.8, \Gamma= 0.5J $ in the 2S-model (dotted line)
and 4S-model (full line)\\

\newpage
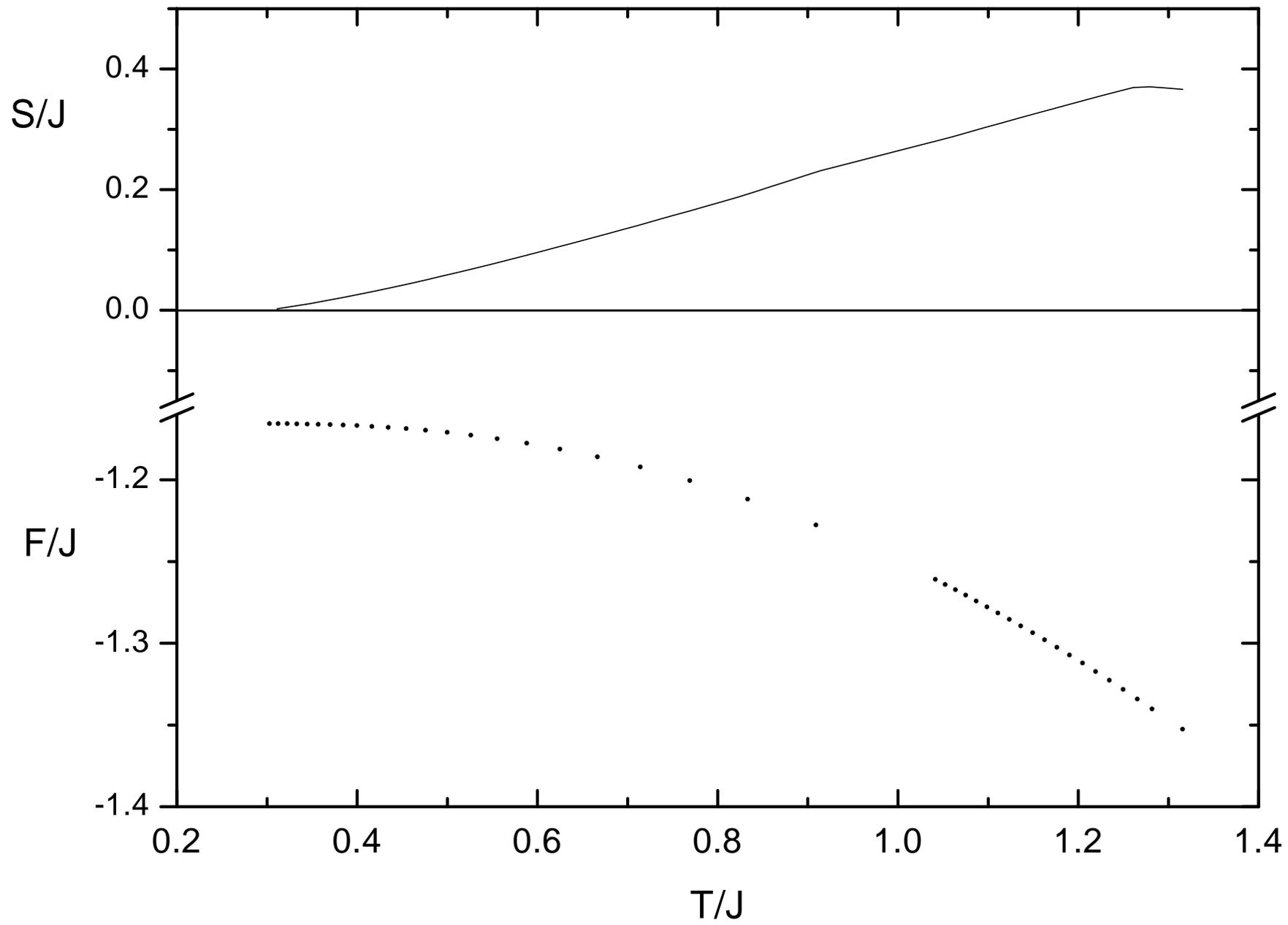
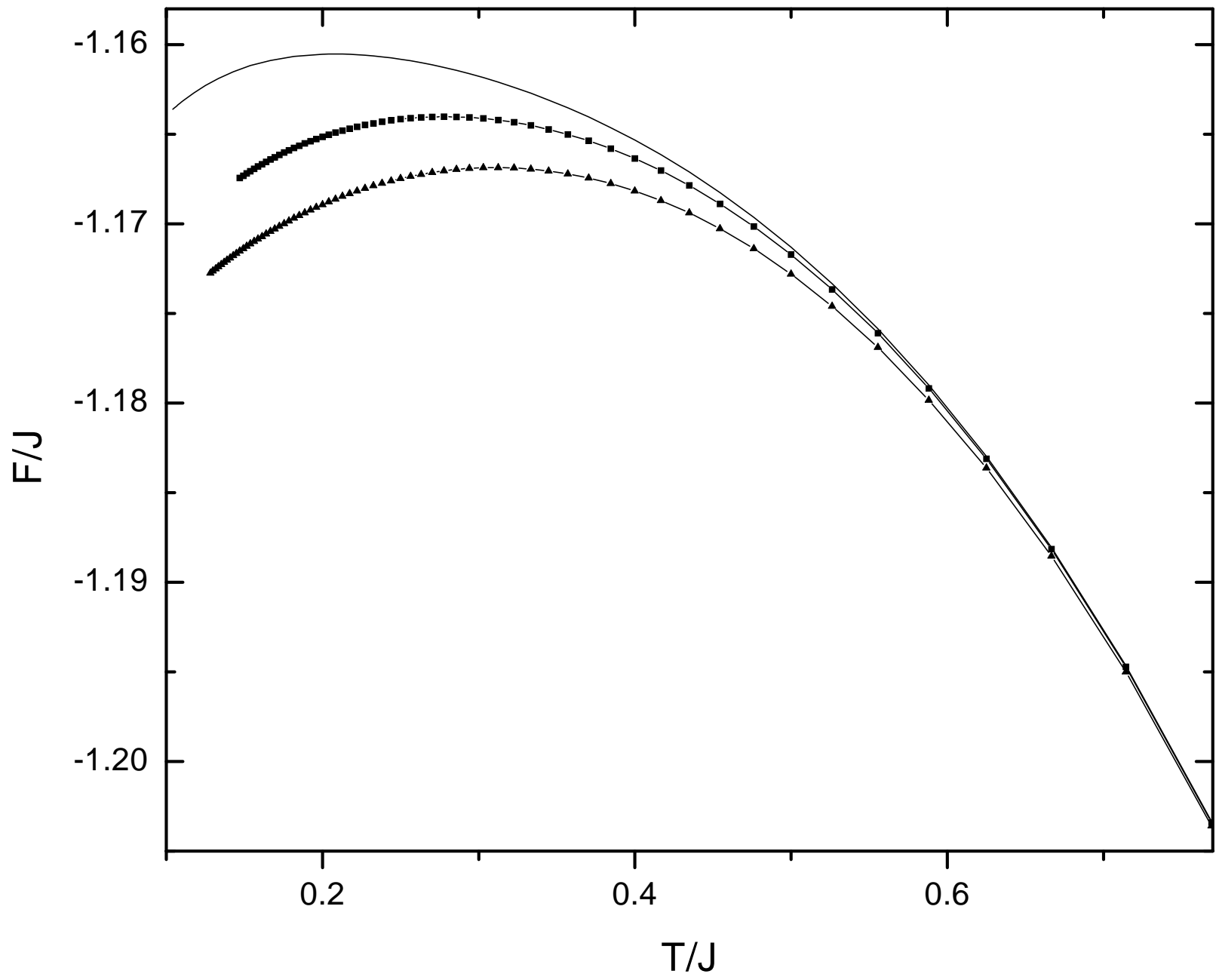
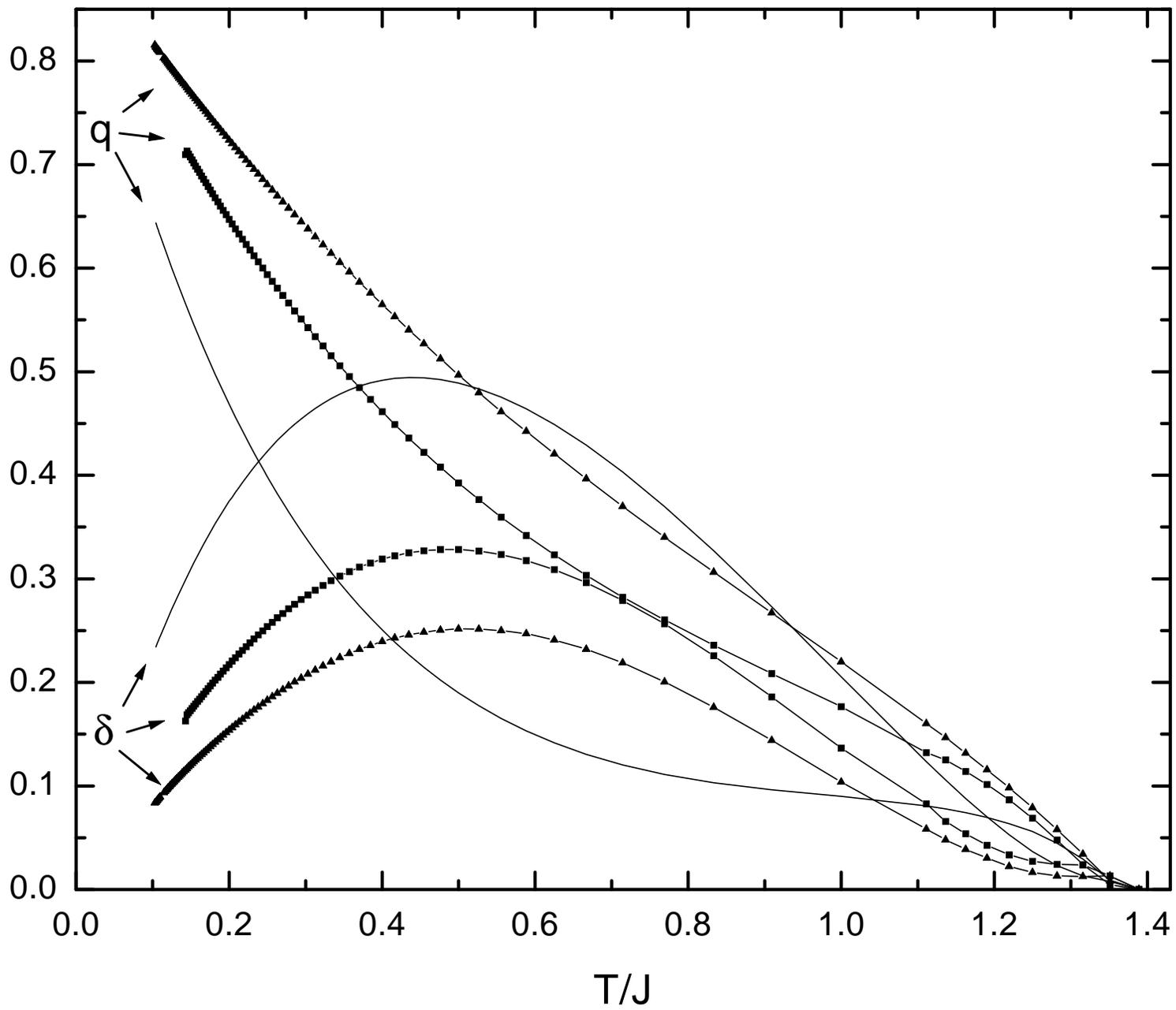
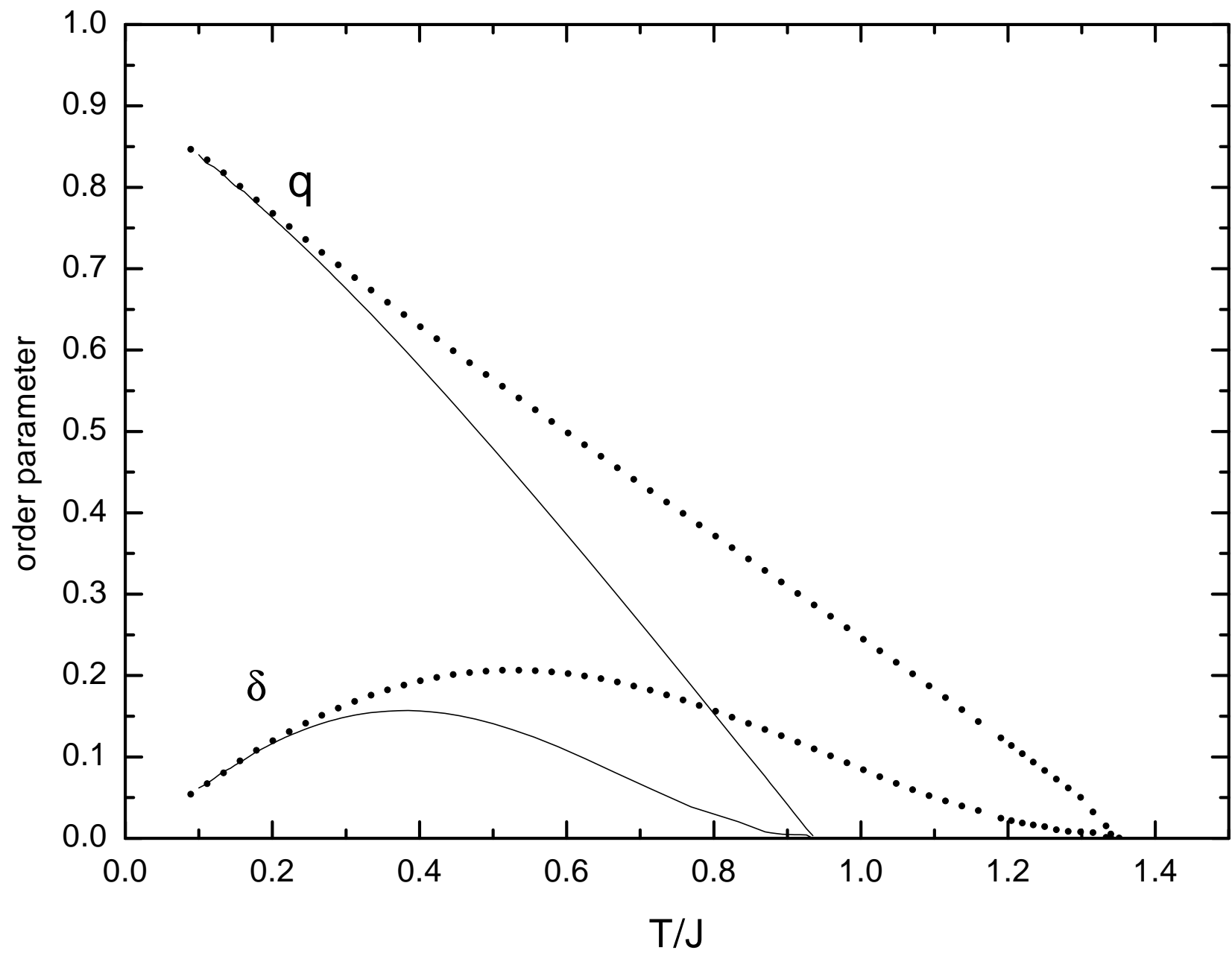
\end{document}